\documentclass[aps,prb,twocolumn,showpacs,floatfix,superscriptaddress,amsfonts,dsfont]{revtex4-1}
\usepackage{graphicx}
\usepackage{dcolumn}
\usepackage{bm}
\usepackage{amsmath,amssymb,amsthm,mathrsfs,amsfonts}
\usepackage{fontenc}
\usepackage{color}
\usepackage{placeins}
\usepackage{array}

\begin{document}


\title{Topological nature of in-gap bound states in disordered large-gap monolayer transition metal dichalcogenides}

\author{Fanyao Qu}
\author{L. Villegas-Lelovsky}
\affiliation{Instituto de F\'isica, Universidade de Brasilia,70910-900, Brasilia, DF, Brazil}
\author{G. S. Diniz}
\affiliation{Instituto de F\'isica, Universidade Federal de Uberl\^andia, Uberl\^andia, MG, Brazil}


\begin{abstract}
We propose a physical model based on disordered (a hole punched inside a material) monolayer transition metal dichalcogenides (TMDs) to demonstrate a large-gap quantum valley Hall insulator. We find an emergence of bound states lying inside the bulk gap of the TMDs. They are strongly affected by spin-valley coupling, rest- and kinetic- mass terms and the hole size. In addition, in the
whole range of the hole size, at least two in-gap bound states with opposite angular momentum, circulating around the edge of the hole, exist. Their topological insulator (TI) feature is analyzed by the Chern number, characterized by spacial distribution of their probabilities and confirmed by energy dispersion curves (Energy vs. angular momentum). It not only sheds light on overcoming low-temperature operating limitation of existing narrow-gap TIs, but also opens an opportunity to realize valley- and spin- qubits.
\end{abstract}

\maketitle

\emph{Introduction} - Transition metal dichalcogenides (TMDs) exhibit diverse states of matter. Depending on their composition, they can be semiconductors (e.g. 2H-$\textrm{MoS}_{2}$, $\textrm{WS}_{2}$) with a band gap in the visible to near-IR frequencies, semi-metals, true metals, and superconductors \cite{Strano}. For certain composition, such as $\textrm{MoS}_{2}$, its electronic structure and optical property can also be tailored by the crystalline structure. For example, 2H-$\textrm{MoS}_{2}$ is semiconducting, whereas 1T-$\textrm{MoS}_{2}$ is metallic. Breaking space-time symmetries in two-dimensional crystals can markedly influence their macroscopic electronic properties. For instance, as the number of stacking layers changes from two to one, crystal inversion symmetry is intrinsically broken, then the band structure of bulk 2H-$\textrm{MoS}_{2}$ transforms from an indirect band gap to a direct one \cite{Strano}. It leads to an enhancement of optical emission intensity and a generation of valley-selective optical polarization, even though the two valleys are energetically degenerate, locked by time reversal symmetry (TRS) \cite{Mak,Jones,Zaumseil,PhysRevLett.113.266804}. On the other hand, as known, perfect crystals do not exist. Hence impurities or defects are inevitably present in the TMDs. They change the geometry or topology of the systems and induce in-gap bound states \cite{Andor}. Then the TMDs with a hole punched inside them may be of the characteristics of large-gap topological insulators (TIs). Up to now, only very few systems has been discovered to possess this property \cite{Xiaofeng}. It obstructs tremendously applications in low-power, multi-functional and high-temperature operating spintronic devices. This motivated us to realize a systematic study on the in-gap bound states in the TMDs.
\begin{figure}[!ht]
\centering
\includegraphics[scale=0.23]{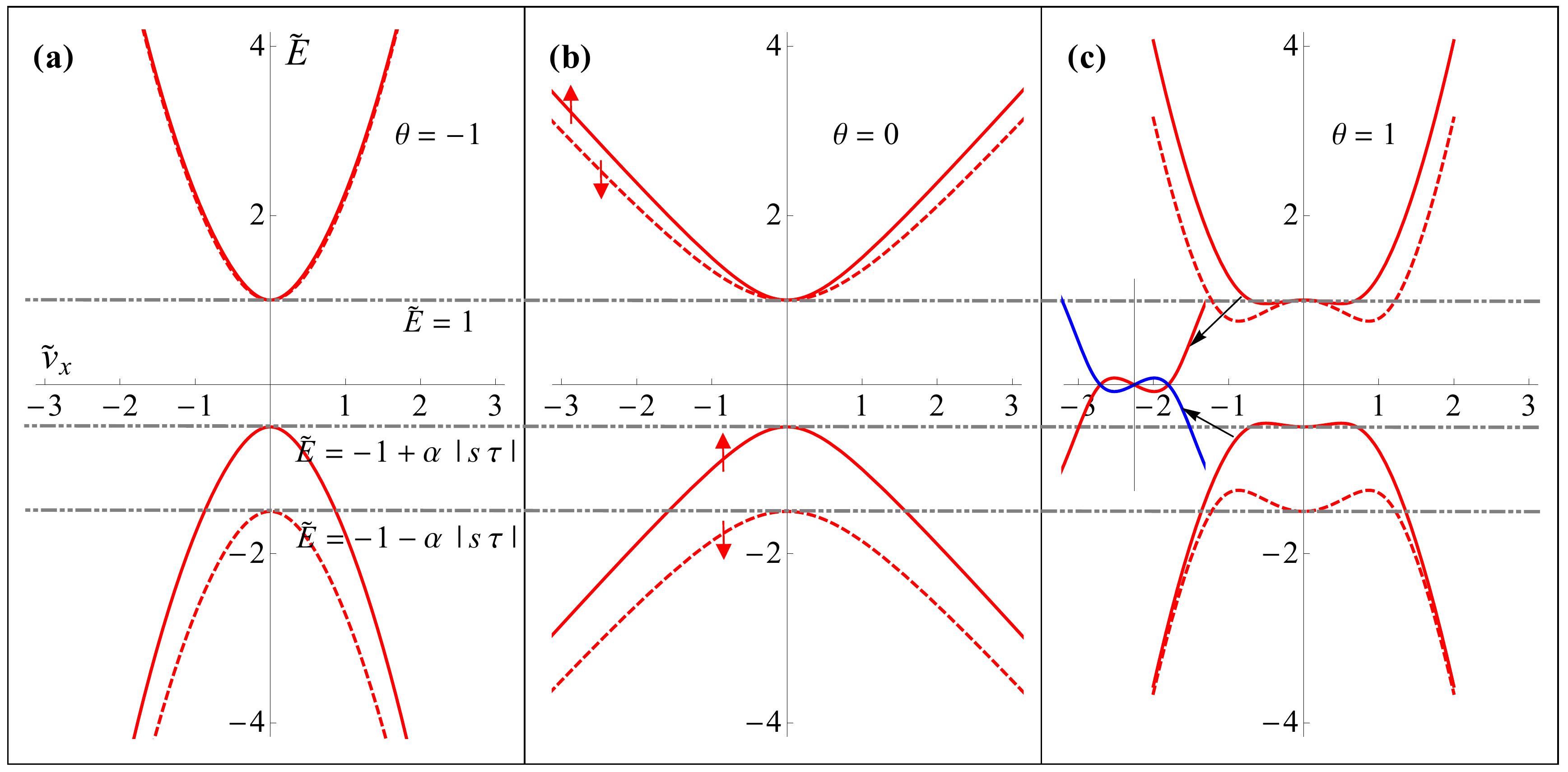}
\caption{(Color online) Energy vs $v_{x}$ of monolayer $\textrm{MoS}_{2}$ for $v_{y} = 0$ and mass related parameter $\theta$ = -1 (a), 0 (b) and 1 (c). Inset in (c) shows derivative of the energy bands (or inflection points) band structure around $K$- or $K^{\prime}$- point for conduction- (red curve) and valence- (blue curve) band.  The vertical arrows indicate the direction of spin in the valley $K$.}
\label{fig1}
\end{figure}

\emph{Theoretical Model} - We start from a description of the low-energy model of monolayer TMDs. This effective two-band model has been derived  from density functional theory calculations \cite{Xiao-1,Xiao-2} and supported by optical experiments \cite{hualing,ArXiv}. In this model, near the Brillouin zone inequivalent corners $K$ ($\tau=1$) and $K^{\prime}$ ($\tau =-1$), the conduction band (CB) arises from $\vert\varphi_{c}\rangle = \vert d_{z^2}\rangle$ orbital with angular momentum $l_{z} =0 $, while the valence band (VB) is approximately from hybridization of $\vert d_{x^{2}-y^{2}}\rangle$ and $\vert d_{xy}\rangle$ orbitals with $l_{z} =2\tau\hbar$, i.e., $\vert\varphi_{v}\rangle$ = $\vert d_{x^{2}-y^{2}}\rangle + i \tau \vert d_{xy}\rangle$. Then in the vicinity of the $K$ and $K^{'}$ valleys, wave functions can be
constructed by using $\vert\varphi_{c}\rangle$ and $\vert\varphi_{v}\rangle$ as basis functions. Therefore, the TMDs can be described by the valley-modified Dirac Hamiltonian,

\begin{equation}
H = H_{0}+ H_{m} +H_{so},
\label{eq1}
\end{equation}
the first term in Eq. \ref{eq1}, $H_{0} = v_{F}(\tau \textbf{p}_x \sigma_x + \textbf{p}_y \sigma_y)$, is the kinetic energy of massless fermions with Fermi velocity $v_{F}$, here $\sigma=(\sigma_{x},\sigma_{y},\sigma_{z})$ are the Pauli matrices in CB- and VB- basis. The second term, $H_{m}=\tau (m v_{F}^2 - \textrm{B} \textbf{p}^2) \sigma_z$, is a mass term. It is composed of two  valley-dependent but spin-independent terms with an opposite sign, where the former is the rest-mass term, determining the bulk gap. While the latter is the quadratic correction in momentum-$\textrm{B} \textbf{p}^2$ to the band gap, presenting parabolicity of the band, here $m$ and $\textrm{B}^{-1}$ have the dimensions of mass. The third term, $H_{so} = (1-\sigma_z) \beta\tau s$, is the strong spin-orbit coupling (SOC) term dependent upon both spin and valley, with $s$ and $\beta$ corresponding to the real spin and the SOC strength (with values ranging from 160-430 meV). For the monolayer TMDs, the energy spectrum is governed by:

\begin{equation}\label{eq8}
E_{\pm}= s \beta \tau \pm \sqrt{(m v_{F}^2 - \textrm{B}p^2 - s \beta \tau)^2 + (p v_{F})^2}.
\end{equation}
Note that the mass term ($H_{m}$) leads to an opening of energy gap ($2mv_{F}^{2}$) between CB and VB at either the $K$ or $K^{\prime}$ valley, but holding the particle-hole symmetry. Notice also that in the absence of the SOC ($\beta=0$), the states $E_{\pm}$ possess SU(4) symmetry associated with the spin and valley degrees of freedom. Thus, there is no topological distinction between particles and holes. However, this is not the case of TMDs, because in these materials $d$-atoms provide strong intrinsic SOCs, especially in the valence band. A combined effect of inversion asymmetry along with strong SOC lifts the degeneracy of spin-up and spin-down states in the same valley. For instance, in the $\tau$- valley, spin splitting is equal to $4 \beta$ in the VB and 0 in the CB. Therefore, both the SU(4) and particle-hole symmetries are broken. However, the valley degeneracy is still preserved by the TRS. It is worth to notice that the SOC mediated band structure is strongly modulated by $\theta= \textrm{B} m$, as shown in Fig. \ref{fig1}. For instance, at either $K$ or $K^{\prime}$ valley, $E_{+}= m v_{F}^2$ and  $E_{-}=- (m v_{F}^2 -2 s \beta \tau) $, respectively. It is worth to comment that for convenience, the following dimensionless quantities $\tilde E =\frac{E}{|m  \mathit{v_{F}^2}|}$; $\theta= \textrm{B} m$; $\alpha=\frac{\beta}{\vert m  \mathit{v_{F}^2}\vert}$ are used in Fig. \ref{fig1} and in the rest of the paper.

The topological properties of the disordered TMDs may also be studied through the Chern number ($\mathcal{C}$) calculation. In order to get the Chern number, let us first calculate the Berry curvature $\Omega_{xy}^{n}(k_{x},k_{y})$ of the n$th$ bands \cite{Xiao-3} using the Kubo formula,
\begin{equation}
\Omega_{xy}^{n}(k_{x},k_{y})=-\sum_{n^{\prime}\neq n}\dfrac{2Im\langle \Psi_{nk}\vert v_{x}\vert \Psi_{n^{\prime}k}\rangle\langle \Psi_{n^{\prime}k}\vert v_{y}\vert \Psi_{nk}\rangle}{(\omega_{n^{\prime}}-\omega_{n})^{2}},
\label{Berry}
\end{equation}
where $\omega_{n}=E_{n}/\hbar$ with $E_{n}$ the energy eigenvalue of the n$th$ band and $v_{x(y)}={\hbar}^{-1}\partial H_{\tau}/\partial k_{x(y)}$ is the Fermi velocity operator. With the Berry curvature at hand, then the $\mathcal{C}$ can be calculated by \cite{Diniz,Marcos}
\begin{equation}
\mathcal{C}=\dfrac{1}{2\pi}\sum_{n}\int_{BZ}d^{2}k \Omega_{xy}^{n},
\label{H4}
\end{equation}
where the summation is taken over the occupied states below the Fermi level and the integration is done over the first Brillouin zone \cite{Diniz,Marcos}. As the Berry curvature is highly peaked around the valleys, thus it is convenient to calculate the Chern number around $K$ and $K^{\prime}$ using $\mathcal{C}=\dfrac{1}{2\pi}\sum_{K,K^{\prime}}\sum_{n=1,2}\int_{-\infty}^{{\infty}}dq_{x}dq_{y} \Omega^{n}_{xy}(q_{x},q_{y})$, where a momentum cutoff is set around each valley for which the $\mathcal{C}$ converges.

\begin{figure}[!ht]
\centering
\includegraphics[scale=1.1]{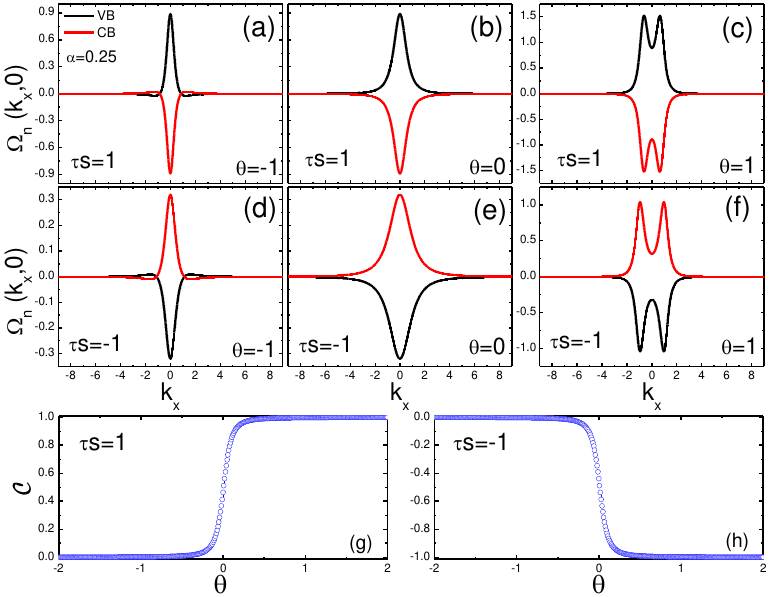}
\caption{(Color online) (a)-(f) Conduction (red curves) and valence (black curves) band Berry curvatures of monolayer $\textrm{MoS}_2$ along $k_{x}$ for different $\theta$ with $s\tau=1$ (upper panel) and $s\tau=-1$ (middle panel). (g)-(h) Calculated Chern number as function of $\theta$. The spin-orbit parameter $\alpha$ is set to 0.25.}
\label{fig2}
\end{figure}

In the pristine TMDs, the Berry curvature is radially symmetric around the valleys, therefore we can restrict the calculation of Berry curvature along one specific direction. Fig. \ref{fig2} shows the Berry curvature for the two energy branches (CB and VB) along $k_x$ for $k_y=0$ with different $\theta$ and $s\tau$. Notice that the states possess the same Berry curvature with its TRS counterpart. While the opposite spin states in the same valley present distinct Berry curvatures. Also it is worth to note that the $\mathcal{C}$ is nonzero for either $s\tau=1$ or $s\tau=-1$ states as $\theta > 0$. In contrast, it is zero in either case of $s\tau$  if $\theta < 0$. As the system is TRS-invariant, the Chern number determined by the sum of the separated contribution of each valley is zero, as one can observe from Fig. \ref{fig2} (g)-(h), where the Chern number as function of $\theta$ is presented. A small applied magnetic field \cite{Andor} or even an intrinsic magnetism induced by magnetic impurities \cite{Cheng,Cong,Andriotis,Cheng2} would be enough to break the TRS, henceforth a nonzero Chern number is expected, which implies a non-vanishing Hall conductivity. This behavior would be of great applicability in the Quantum Hall studies \cite{Xiao-2,Xiaofeng,Ma,Sodemann}.

\begin{figure}[!ht]
\centering
\includegraphics[scale=1.0]{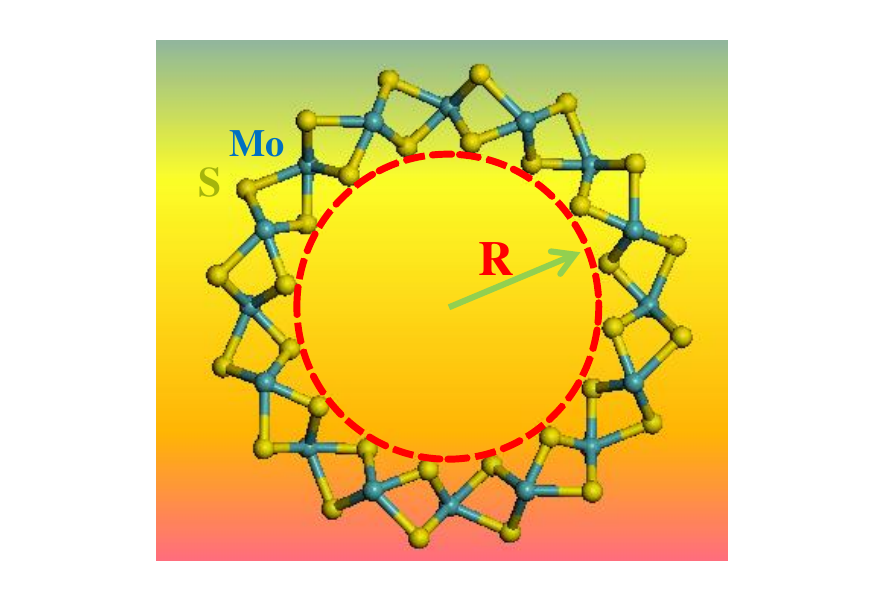}
\caption{(Color online) Schematic diagram of a vacancy in monolayer $\textrm{MoS}_2$. A hole with a radius of $R$ being punched from monolayer $\textrm{MoS}_{2}$ represents the vacancy. $\textrm{S}$ and $\textrm{Mo}$ stand for the sulphur and molybdenum atoms, respectively}
\label{fig3}
\end{figure}

\emph{The vacancy model} - The disordered system studied in this work is created by punching a hole of radius $R$ in the center of the two-dimensional TMDs, as schematically shown in Fig. \ref{fig3}. Because of the rotational symmetry of the system, the polar coordinates ($\tilde{r}, \theta$) are utilized. As done in the previous section, let us introduce dimensionless quantity $\tilde r=\frac{r}{\xi}$, where $\xi^{-1}=\frac{\vert m \mathit{v}\vert}{\hbar}$ denotes the characteristic wavelength of the system. In this notation, the 2x2 spin-valley coupled Hamiltonian $H$ is defined by $h_{11}=1+\theta \left(-\frac{ \ell ^2}{\tilde{r}^2}+\frac{\partial^2 }{\partial \tilde{r}^2}+\frac{1}{\tilde{\tilde{r}}}\frac{\partial }{\partial \tilde{r}}\right) -\tilde{E}$, $h_{12}=-i
    \left(\tau  \frac{\partial }{\partial \tilde{r}}+\frac{\ell +\tau}{\tilde{r}}\right)$, $h_{21}=-i \left(\tau  \frac{\partial }{\partial \tilde{r}}-\frac{\ell  }{\tilde{r}}\right)$ and $h_{22}=-1-\theta \left(-\frac{ (\ell +\tau )^2}{\tilde{r}^2}+\frac{\partial^2 }{\partial \tilde{r}^2}+\frac{1}{\tilde{r}}\frac{\partial }{\partial \tilde{r}}\right)-\tilde{E} +\alpha s \tau$. Then eigenstates of this modified Dirac Hamiltonian are constituted by two spinor radial components
\begin{equation}\label{eq3}
  \psi(\tilde{r})=\begin{pmatrix}c_{\ell}K_{\ell}(\lambda \tilde{r}) \\ d_{\ell}K_{\ell+\tau}(\lambda \tilde{r})\end{pmatrix},
\end{equation}
where $\ell=j-1/2$, $l$ and $j$ are the $z$-component of orbital and total angular momentum, $K_{n}(x)$ is the modified Bessel functions of second kind, $c_{\ell}$ and $d_{\ell}$ are expansion coefficients. Using hard wall boundary conditions $\psi(R/\xi)= 0$ and $\psi(\infty)= 0$, we arrive at the transcendental equation for the bound state energies,

\begin{align}\label{eq6}
\frac{\left(\lambda _1^2+\frac{1-\tilde{E}}{\theta}\right) K_{\ell +\tau }\left(R \lambda
   _1\right)}{\lambda _1 K_{\ell }\left(R \lambda _1\right)}=\frac{\left(\lambda
   _2^2+\frac{1-\tilde{E}}{\theta }\right) K_{\ell +\tau }\left(R \lambda _2\right)}{\lambda
   _2 K_{\ell }\left(R \lambda _2\right)},
\end{align}
with $\lambda_{1,2}^2= \left[1-\theta  (2-\alpha  s \tau )\pm \gamma(s,\alpha,\theta,\tau)\right]/2 \theta^2$, and $\gamma(s,\alpha,\theta,\tau)=\sqrt{1-2 \theta  (2-\alpha  s \tau)+\theta ^2  (2 \tilde{E}-\alpha  s \tau )^2}$.

\begin{figure}[!ht]
\centering
\includegraphics[scale=0.3]{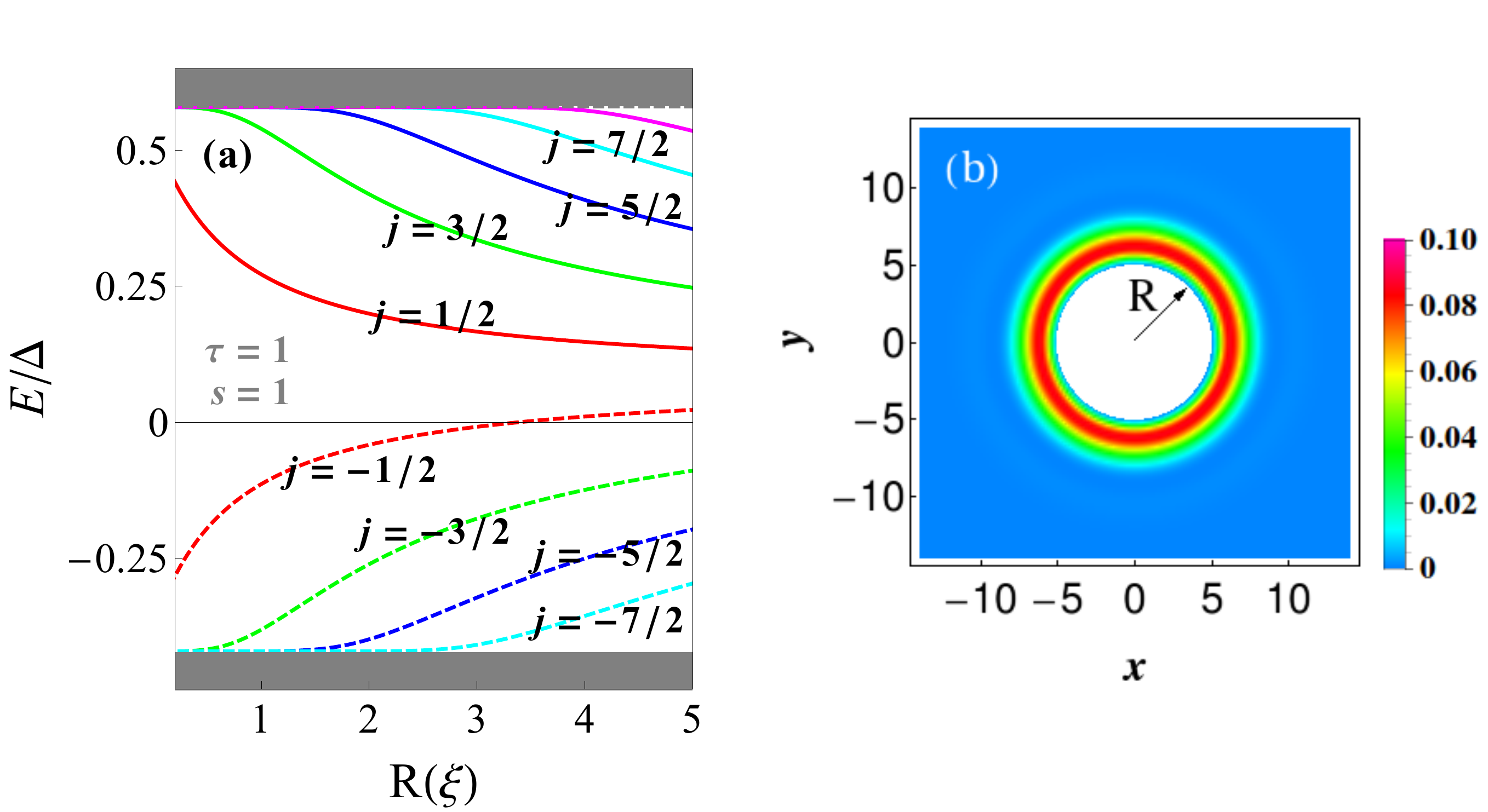}
\caption{(Color online)(a) Energies (in units of the band gap $\Delta$) of in-gap bound states of disordered monolayer TMDs as a function of vacancy radius $R$ for $\tau=1$, $s=1$, $\theta=1$ and $\alpha=0.25$. (b) Probability distribution of the $j = 1/2$ states in the monolayer TMDs with a vacancy $R=5$. $j$ is the $z$-component of the total angular momentum. In (a) gray and light-gray areas stand for the bulk bands, and horizontal black line marks the Fermi energy level.}
\label{fig4}
\end{figure}

Fig. \ref{fig4}(a) illustrates in-gap bound states as a function of the hole radius for the disordered monolayer TMDs. Notice that for a macroscopically large $R$, several bound states around the cavity, determined by a numerical solution of the energy spectrum from Eq. \ref{eq6}, exist. For $R= 5$, for example, there are nine in-gap bound states. The feature of theses states are clearly characterized by edge states, as demonstrated in Fig. 4(b) for $j = 1/2$ state. In addition, the particle-hole symmetry is a broken within a valley. By shrinking $R$, the energy separation of these edge states increases, and the edge states with higher $j$ are pushed away of the energy gap gradually due to an enhancement of the quantum confinement induced by the boundary of the vacancy. When the vacancy radius is reduced up to several angstroms, only two in-gap bound states is observed with $\vert j\vert=1/2$, which are governed by

\begin{equation}\label{eq8}
\tilde E_\pm= \frac{\alpha  s \tau }{2}\pm \tilde\Delta,
\end{equation}
can survive, here we assume that \cite{limit}
\begin{eqnarray}\label{eq9}
\tilde\Delta=\left\{
\begin{array}{ll}
2 {\rm ,\ }& 0<\theta <\frac{1}{2-\alpha  s \tau } \\\\
\frac{\sqrt{ (\theta  (4-2 \alpha  s \tau )-1)}}{|\theta|}{\rm ,\ }& \theta >\frac{1}{2-\alpha  s \tau }.
\end{array}
\right.
\end{eqnarray}

From Eq. (\ref{eq8}) and (\ref{eq9}), note that the in-gap bound states, i.e., the phase of quantum valley insulator in monolayer TMDs, exists provided that $\theta \geq \frac{1}{4-2 \alpha s \tau }$. Note also that this condition is determined only by the physical parameters of the TMDs rather than a radius of the hole.
\begin{figure}[!ht]
\centering
\includegraphics[scale=0.4]{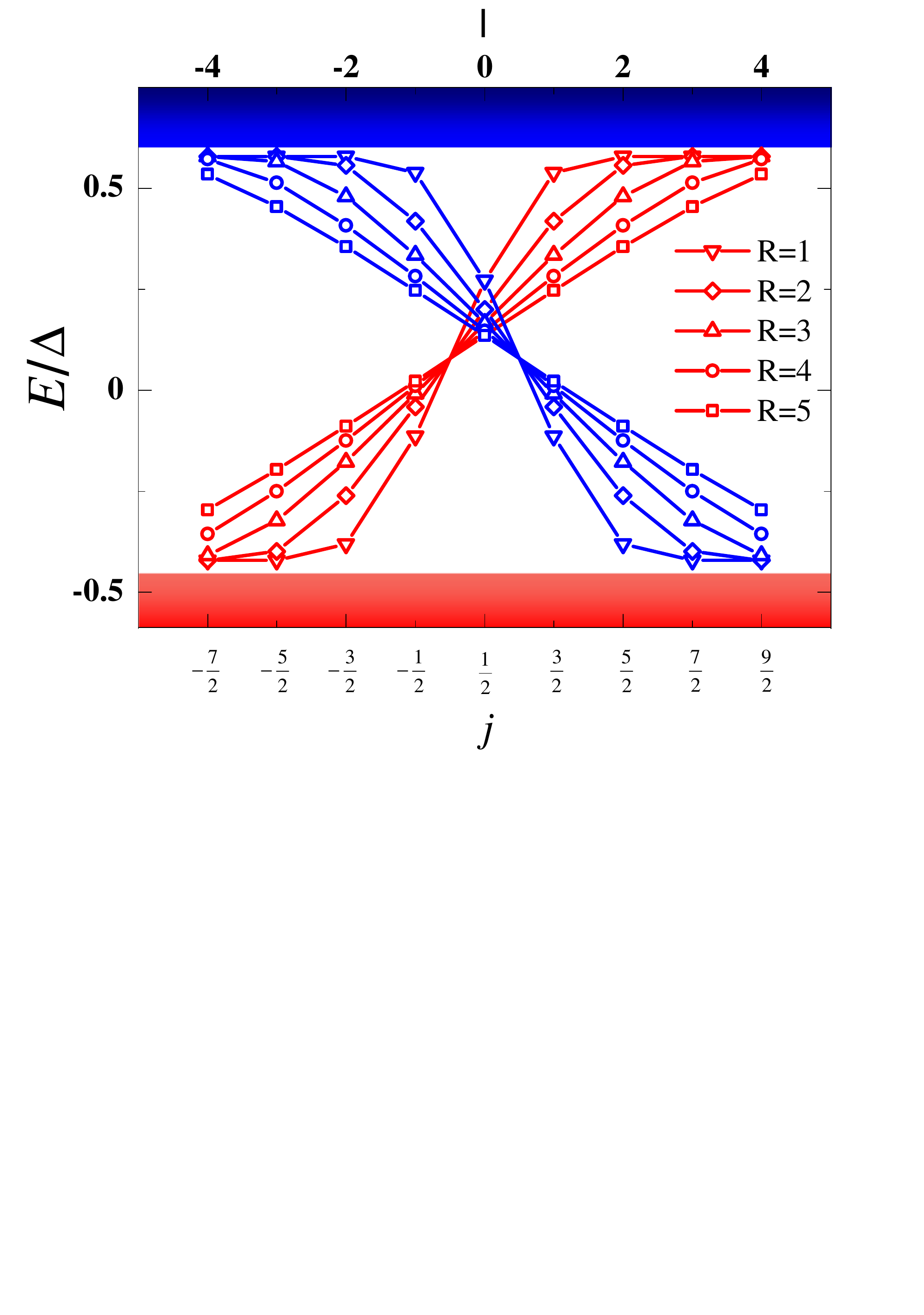}
\caption{(Color online) Discrete energy spectrum as a function of $z$-component of total (orbital) angular momentum quantum number $j$ ($l$) for $\theta=1$, $\alpha=0.25$ and five different vacancy radius $R$. The red symbols represent the energies of the states characterized by $\tau=1$ and $s=1$, while the blue ones correspond to $\tau=-1$ and $s=-1$. The lines are just a guide for eyes.}
\label{fig5}
\end{figure}

Fig. \ref{fig5} shows the energy spectrum of in-gap bound states in valley $K$ with spin-up (red symbols) and in valley $K^{\prime}$ with spin-down (blue symbols). Notice that for a given valley, say valley $K$, the states with opposite total angular momentum but the same spin show a particle-hole symmetry breaking ($E(j) \neq - E(-j)$). Whereas in the combined spectrum of two valleys, the valley degeneracy of the states with opposite spin and orbital angular momentum is restored. Besides, electrons with opposite spin and orbital angular momentum in $K$ and $K^{\prime}$ valleys propagate in opposite directions, manifested in the opposite sign of the inclination of the correspondent energy curves. Thus, these pairs of states form paired helical edge states, which are in good agreement with the edge-state solutions in the two-dimensional system if we take $k = (l+1/2)/R$ for a large $R$. It is worth to comment that the presence of the bound states are not sensitive to the shape of the vacancy due to their topological origin. Hence the results presented here are general, which are independent on the geometric form of the vacancy.

To reveal the underlying physics of topological features of in-gap bound states in disordered TMDs, let us analyze the Fig. \ref{fig4} and  \ref{fig5} from symmetry point of view. To do so, first, we recall four useful symmetry operators: (1) Inversion symmetry operation: interchanges the valleys and inverts the sign of both $\textbf{r}$ and $\textbf{p}$, i.e., $P$($\textbf{r}$, $\textbf{p}$, $\tau$, s)$\rightarrow$ (-$\textbf{r}$, -$\textbf{p}$, -$\tau$, s). This symmetry operation can be properly described by an operator $P = \tau_x \otimes \sigma_x $; (2) Effective inversion symmetry operation: similar with $P$, but the symmetry operation is realized only within the same valley, i.e., $P_{e}$($\textbf{r}$, $\textbf{p}$, $\tau$, s)$\rightarrow$ (-$\textbf{r}$, -$\textbf{p}$, $\tau$, s), correspondent operator given by $P_{e} = I_\tau \otimes \sigma_x$, here $I_\tau$ is identity matrix; (3) TRS operation: not only inverts the sign of $s$, $\textbf{p}$ and $\textbf{l}$, but also interchanges the valleys, i.e., $T(\textbf{r}, \textbf{p}, $\textbf{l}$, \tau, s)\rightarrow (\textbf{r}, -\textbf{p}, -$\textbf{l}$, -\tau, -s)$. Then the correspondent operator reads $T = i \tau_x \otimes s_{y}C $, where $C$ is the complex conjugate operator; (4) Effective TRS operation: like $T$, but the symmetry operation is only realized within the same valley, i.e., $T_{e}(\textbf{r}, \textbf{p}, $\textbf{l}$, \tau, s)\rightarrow (\textbf{r}, -\textbf{p}, -$\textbf{l}$, \tau, -s)$, the correspondent symmetry operator described by $T_{e} = I_\tau \otimes s_{y} C $. Therefore, $P_{e}$ and $T_{e}$ act like inversion symmetry and TRS operators in a single valley, respectively. By using these properties of symmetry operators, we can analyze the results shown in Fig. \ref{fig4} and  \ref{fig5}. Although the Hamiltonian is invariant under the action of $P$, it does not commute with $P_{e}$. Thus, an effective inversion symmetry is broken. Consequently, the states with opposite angular momentum in the same valley possess different energies. On the other hand, the first term in $H_{SO}$ does not commute with neither $P_{e}$ nor $T_{e}$. Hence the states with the same spin but opposite angular momentum in the different valleys are asymmetric, and no particle-hole symmetry exists in a single valley. However, the total Hamiltonian and boundary condition preserves the TRS. As expected, the Fig. \ref{fig5} illustrates  $E(j,s,\tau) = E(-j,-s,-\tau)$.

\emph{Conclusion} - In summary, we have studied the general conditions for the in-gap bound states formation in a TMD. We find that the combined effects of hole border confinement, SOC, rest- and kinetic mass terms strongly change energy spectrum of these bound states. However, under certain conditions, at least two-bound states exist at the whole range of hole-size. The topological nature of these in-gap bound states manifest itself in their probability distribution functions along with the sign of the Fermi velocity of two TRS states in different valley. The energy spectra and topological feature are interpreted by means of symmetry argument analysis.

\emph{\textbf{Acknowledgements}--}We acknowledge financial support from FAPEMIG, FAP-DF, CAPES and CNPq.

\vspace{-1.5em}

%

\end{document}